\documentclass[12pt]{iopart}

\usepackage{hyperref}
\hypersetup{
  colorlinks,%
  citecolor=black,%
  filecolor=black,%
  linkcolor=black,%
  urlcolor=blue
  }

\usepackage{multirow}
\usepackage{float}
\usepackage[style=numeric-comp, sorting=none, backend=bibtex]{biblatex}
\usepackage{wrapfig}
\usepackage{orcidlink}

\usepackage{lineno}

\newcommand{\orcidA}{\orcidlink{0000-0003-2849-0120}} 
\newcommand{\orcidB}{\orcidlink{0000-0001-9223-6480}} 

\begin{document}

\title[ML based KNO-scaling of charged hadron multiplicities with {\sc Hijing++}]{Machine Learning based KNO-scaling of charged hadron multiplicities with {\sc Hijing++}}

\author{Gábor Bíró$^{1,2}$\orcidA{}, Gergely Gábor Barnaföldi$^1$\orcidB{}}
   
\address{$^1$Wigner Research Center for Physics, 29--33 Konkoly--Thege Mikl\'os Str., H-1121 Budapest, Hungary.}
\address{$^2$Institute of Physics, E\"otv\"os Lor\'and University, 1/A P\'azm\'any P\'eter S\'et\'any, H-1117  Budapest, Hungary.}

\ead{biro.gabor@wigner.hu; barnafoldi.gergely@wigner.hu}

\begin{indented}
  \item[]\today
  \end{indented}

\begin{abstract}
The scaling properties of the final state charged hadron and mean jet multiplicity distributions, calculated by deep residual neural network architectures with different complexities are presented. The parton-level input of the neural networks are generated by the \textsc{Hijing++} Monte Carlo event generator. Hadronization neural networks, trained with $\sqrt{s}=7$~TeV events are utilized to perform predictions for various LHC energies from $\sqrt{s}=0.9$~TeV to 13~TeV. KNO-scaling properties were adopted by the networks at hadronic level.
\end{abstract}

%
%
%
%
%

\section{Introduction}







Modern developments in Machine Learning methods led us to use these techniques in the field of high-energy physics (HEP) with great benefits~\cite{Feickert:2021ajf, chollet2015keras, abadi2016tensorflow, Mallick:2022alr, Monk:2018zsb}. Applications of the artificial intelligence hopefully not only provide solution for so far unsolved questions, but may help to improve physical models by recognizing and investigating the inner correlations from these new approaches. In our recent works, Deep Neural Networks (DNN) were proposed to calculate the hadron-level statistical properties of collision events from the parton-level input, which was pre-calculated and trained by the widely used \textsc{Pythia 8} Monte Carlo (MC) event generator~\cite{Biro:2021zgm, Sjostrand:1982fn, Sjostrand:2014zea, he2015deep, Biro:2022zhl}. We showed that the application of relatively simple neural network models preserve the strong KNO-scaling of the hadronic final-state production yields and their multiplicity distributions at energies available at the Large Hadron Collider (LHC). Another observation was that, despite the models were trained exclusively at one fixed center-of-mass energy of $\sqrt{s}=7$ TeV~\cite{Biro:2022zhl, KOBA1972317, Hegyi:2000sp, Vertesi:2020utz}, the acquired scaling properties result in the application of the same network in more general kinematical ranges.

\begin{figure}[htb]
\centering
\hfill
\includegraphics[width=0.85\linewidth]{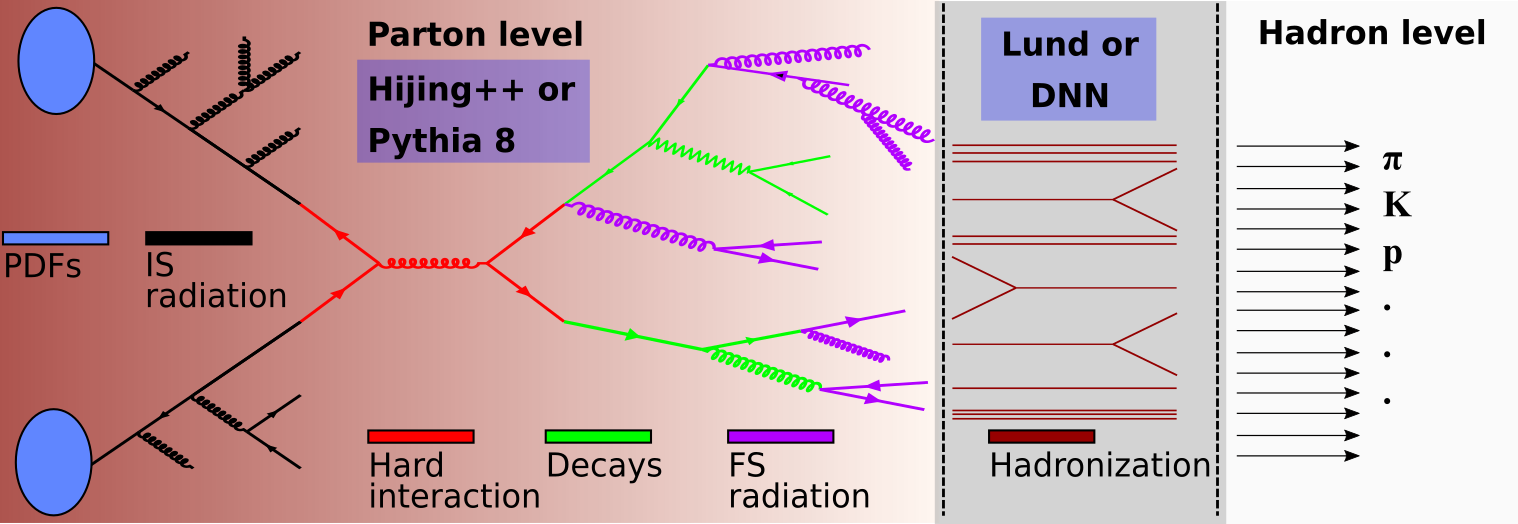}
\hfill
\caption{The schematic overview of the investigated processes and cross-checks.}
\label{fig:blackbox}
\end{figure}

The \textsc{Hijing++} (Heavy Ion Jet INteraction Generator, C++ version) is the new generation of the popular Monte Carlo event generator for heavy-ion physics. This program code is under final tests in the development timeline, and the latest, tuned version is already performing well with data~\cite{Biro:2019ngd, Biro:2019ijx, Biro:2018ntr}. In the current study, the previously proposed and {\sc Pythia 8}-trained, ML-based hadronization models were used to investigate KNO-like scaling behaviour. In order to test more inclusively the neural network (NN) model, the parton-level input of the ML-based hadronization model were generated by the \textsc{Hijing++} in this study. Replacing the Lund hadronization model with a DNN-based one can provide a cross-check and valuable input for the validation of the hadronization model, as it is presented in Fig.~\ref{fig:blackbox}.





\section{The applied models}



One of the focus of our interest is to investigate, whether a neural network is able to represent the properties of the hadronization, especially at the non-perturbative regime of quantum chromodynamics. Within this kinematical regime the physical description lacks of first principle calculations, therefore only complex phenomenological models exist with large sets of inner parameters.
Since it has been proven that a complex deep neural network can pick up the properties of the jet evolution~\cite{Monk:2018zsb}, indeed presenting QCD-like scaling properties~\cite{Biro:2021zgm}, a machine learning-based hadronization model is well motivated. On the other hand, the universality of our hadronization network module is a key concept for the further developments, therefore we investigated the ML-based model by inserting it to an another Monte Carlo generator framework for cross-check tests. 

The DNN hadronization models were developed by applying the popular {\sc ResNet} architecture~\cite{he2015deep, chollet2015keras, abadi2016tensorflow}. Two models with different complexities were proposed, designated as 'Model L' and 'Model S', with respect to the size of the hyperparameter space. Using DNN-based hadronization models in the original {\sc Pythia 8}-based Monte Carlo framework we were able to reproduce the measured charged hadron multiplicity distribution, jet-multiplicity distribution, and observables {\it vs.} event activity classifiers in 'jetty' events (i.e. events where at least 2 jets with $p_{T_J}\geq40$ GeV and $R=0.4$ are present) within a wide range of LHC energies. These correlated well with the physical expectations.

The \textsc{Hijing++} model is the successor of the original {\sc Fortran Hijing}, completely rewritten in modern C++ programming language~\cite{GYULASSY1994307, Biro:2018ntr, Biro:2019ijx}. 
The core concepts of its physics engine are the well-known wounded nucleon model with energy-dependent minijet production~\cite{Wang19913501}, with the Lund string fragmentation model of \textsc{Pythia 8}, taking care of the hadronization~\cite{Sjostrand:1982fn,Sjostrand:2014zea,Cacciari:2011ma}. It has new, modern built-in features such as modularity and CPU multithreading, whereas the underlying physics has been revamped and tuned for the RHIC-LHC energy era. 

In this current study, our concept follows the idea presented in Fig.~\ref{fig:blackbox}: the partonic initial state events, that are the inputs of the DNN-hadronization models, are calculated with \textsc{Hijing++} with the same event criteria that has been applied in Refs.~\cite{Biro:2021zgm, Biro:2022zhl}. Observables are presented and compared to the original Monte Carlo generated events.


\section{Results}

\begin{figure}[htb]
\centering
\hfill
\includegraphics[width=0.49\linewidth]{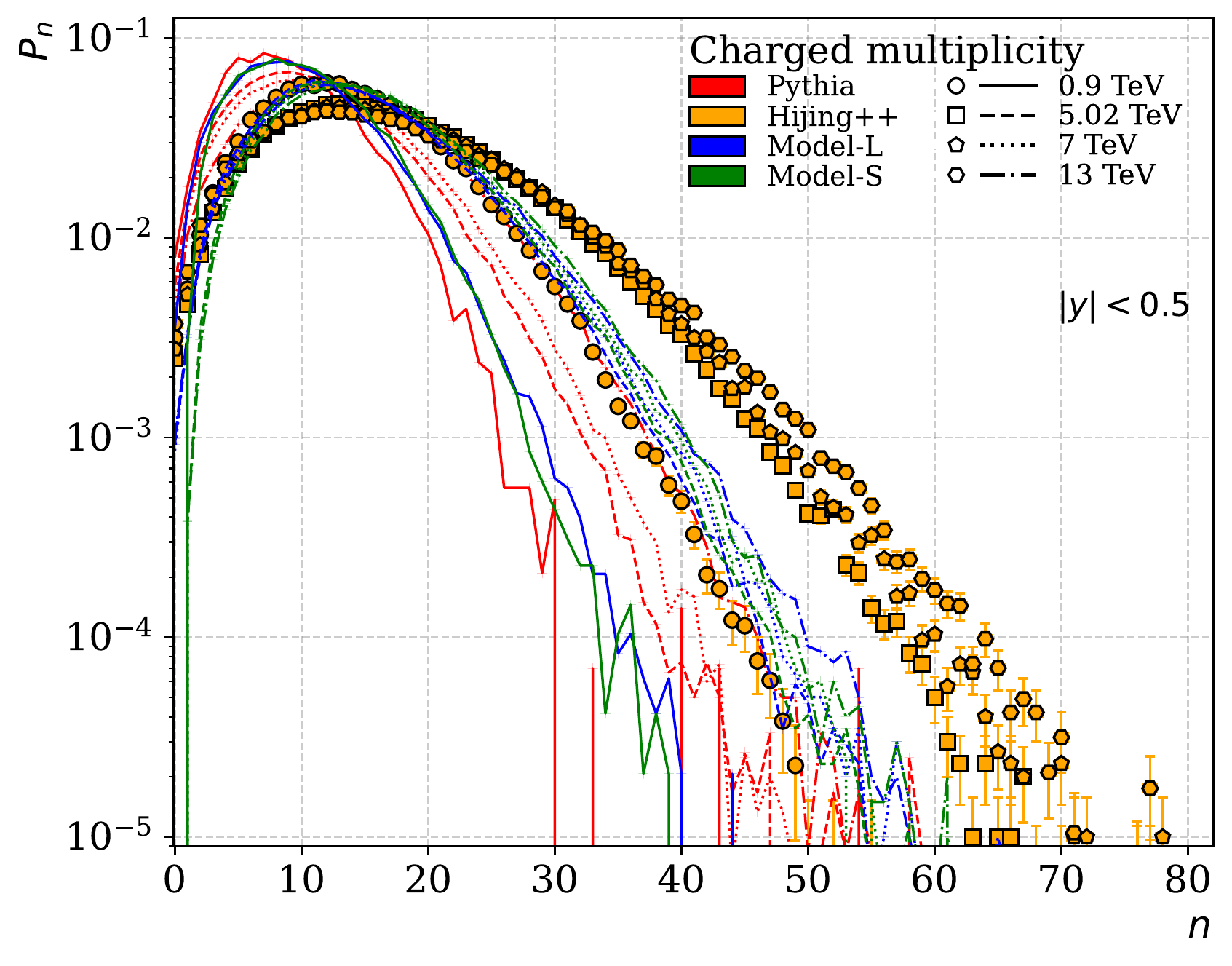}
\hfill
\includegraphics[width=0.49\linewidth]{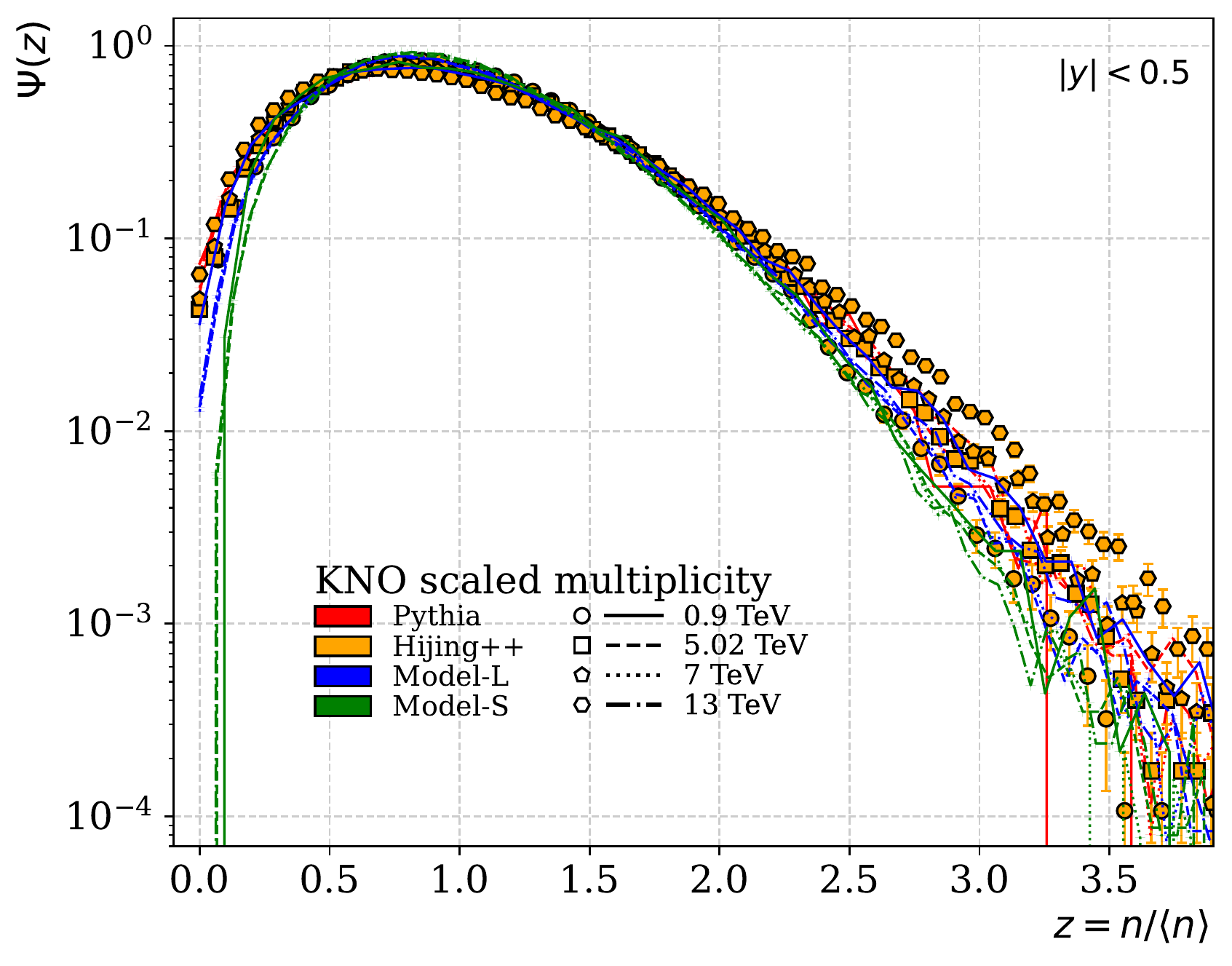}
\hfill
\caption{Mid-rapidity multiplicity (\textit{left panel}) and KNO-scaled distributions (\textit{right panel}) of charged hadrons in proton-proton collisions at LHC energies.}
\label{fig:mult1}
\end{figure}

The multiplicity distributions of charged hadrons stemming from proton-proton collisions are presented on the {\sl left panel} of Fig.~\ref{fig:mult1}, where the \textsc{Hijing++}-calculated values (orange markers) are compared with the NN-predicted results for various c.m. energies in mid-rapidity, $|y|<0.5$ (blue and green lines). The original \textsc{Pythia 8} results are also shown with red lines for reference. Each curve has 300k generated events. The {\sl right panel} of Fig.~\ref{fig:mult1} shows the corresponding $P_n=\frac{1}{\left<n\right>}\Psi\left(\frac{n}{\left<n\right>}\right)$ scaling functions, with the joint curves presenting the effect of KNO-like scaling. 


The primer observation on the multiplicity distribution is that, the \textsc{Hijing++} results display deviation compared to \textsc{Pythia 8} ones. A significant excess contribution appears at higher multiplicity classes against the low-multiplicity region. This difference is not surprising, since the phenomenological mechanisms of the two models are different in the non-perturbative regime. Minijet production in {\sc Hijing++} generates more hadrons in the mid-multiplicity range. The predictions from the NN-based Model~L and Model~S lie between the two Monte Carlo models. Recalling that, NN-based model were trained with the fixed 7~TeV c.m. energy in {\sc Pythia~8} data only, the presented multiplicity distributions of these networks convoluted with {\sc Hijing++} is closer to the original {\sc Pythia~8}-calculated curves at all energies. Indeed, the trends are more {\sc Pythia~8}-like. This supports the idea, that hadronization plays a more significant role in the multiplicity production, than the parton shower evolution---which latter is different in the two Monte Carlo generators. 

We investigated, whether KNO-scaling of the multiplicities is preserved. One can observe on the {\sl right panel} of Fig.~\ref{fig:mult1} that, after applying the KNO-transformation, good agreement were found among all the models up to the highest multiplicities. It is also true for all datasets, that the larger the multiplicity, the stronger the violation of the KNO-scaling is, which has been showed experimentally as well~\cite{CMS:2010qvf}. This violation here were found to be more remarkable for the original \textsc{Hijing++} data. Model L and Model S scale well in parallel to these, apart form the lowest multiplicity values, where the applied cuts and statistics limit the training process. 

\begin{figure}[htb]
\centering
\hfill
\includegraphics[width=0.49\linewidth]{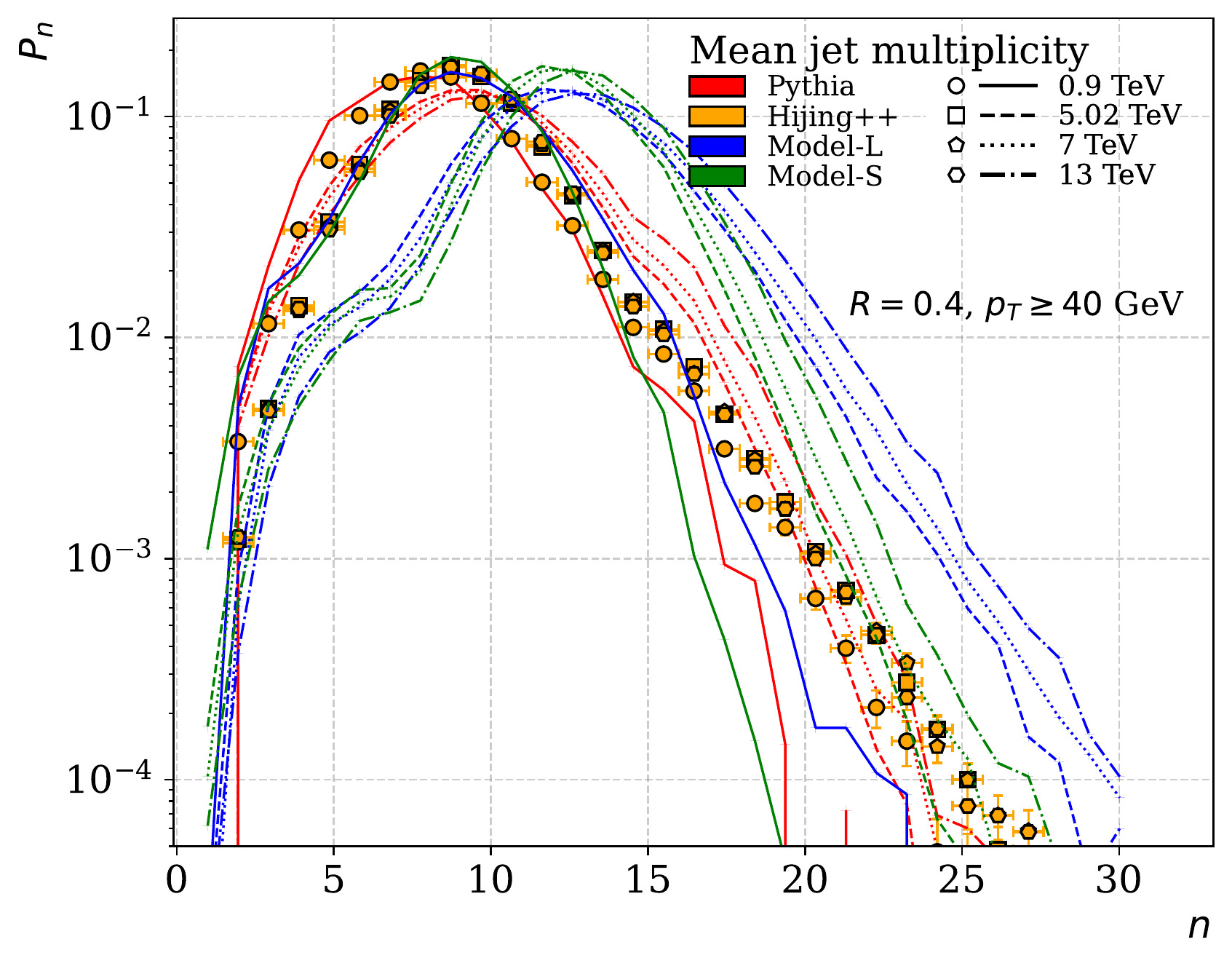}
\hfill
\includegraphics[width=0.49\linewidth]{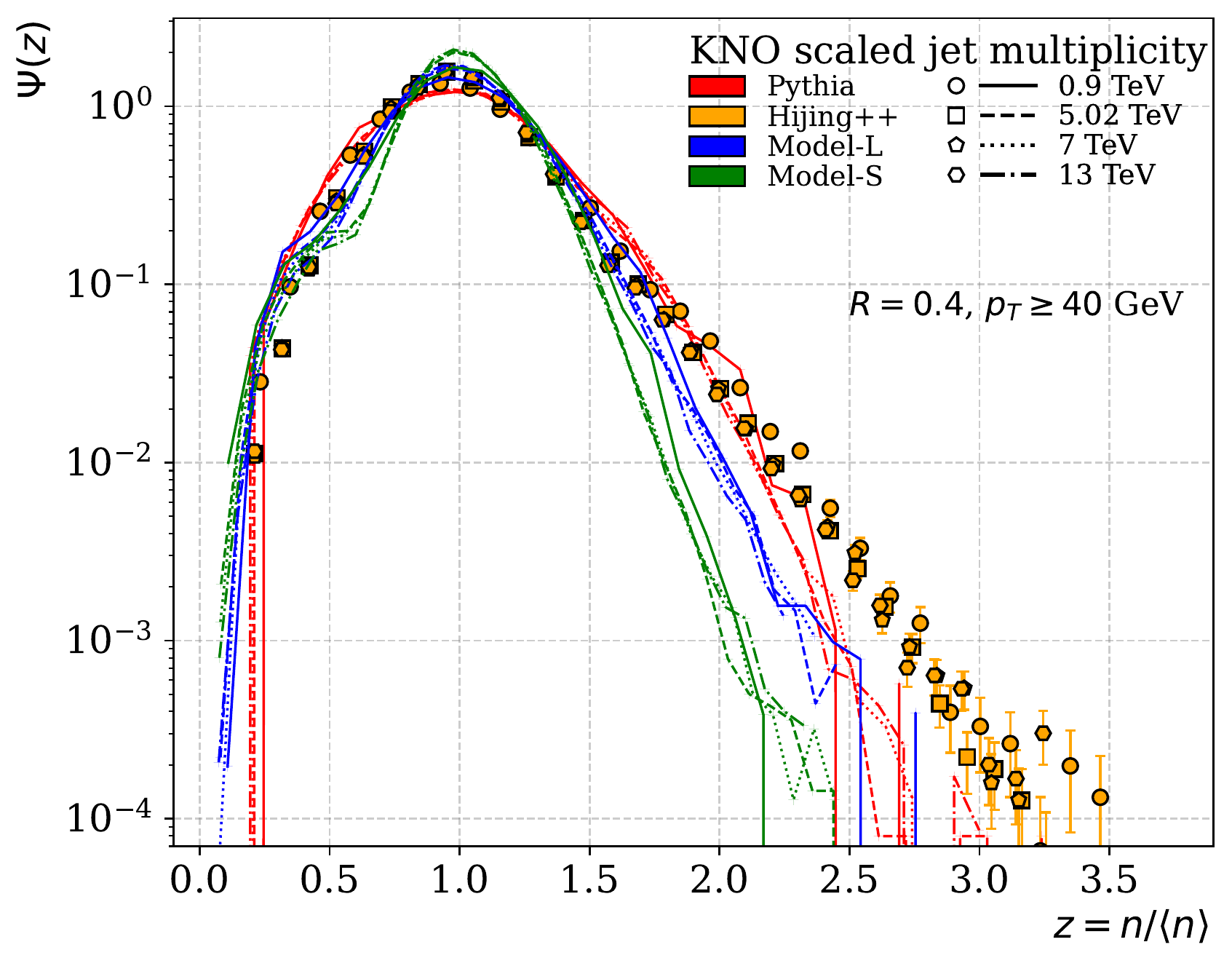}
\hfill
\caption{Jet mean multiplicity (\textit{left panel}) and KNO-scaled jet mean multiplicity (\textit{right panel}) in proton-proton collisions, for jets with $p_{T_J}\geq40$ GeV and $R=0.4$.}
\label{fig:mult2}
\end{figure}

The mean jet multiplicity distributions and the their KNO-scaled curves are shown in the {\sl left} and {\sl right} panels of Figure~\ref{fig:mult2}, respectively. The deviation between the distribution shapes of the original Monte Carlo model results mostly vanishes. In this high-momentum fragmentation regime minor impact form the soft non-perturbative sector is present, therefore difference between the MC calculations appears only at the highest multiplicity values. 

In contrary to the above agreement, the NN-based Model~L and Model~S present irregular, double-bump structure in the mean jet multiplicity distributions. The magnitude of this effect is independent of the hyperparameter-space volume but getting stronger for higher $\sqrt{s}$ values. 
This suggests that though the global jet structure (e.g. the mean jet multiplicity) is similar among the two MC models, the sub-structure is quite different. This effect requires further investigations on a per-jet basis. 

The KNO-scaled mean jet multiplicities on the {\sl right} panel of Fig.~\ref{fig:mult2} are similar to the results that we have seen previously: curve shapes are mostly universal at all energies for each investigated model---again with good agreement between the Monte Carlo models, apart form the highest multiplicity bins, which lack of statistics. The shapes of the scaled distributions are different for the NN-based Model~L and Model~S, separated to two branches by the size of the hyperparameter space volume.

\section{Summary}
 
In this contribution the scaling properties of charged hadron multiplicities and jets at LHC energies, stemming from proton-proton collisions were presented. The multiplicity distributions were determined by two Monte Carlo event generators and deep neural network based hadronization models. The neural network results presented a KNO-scaling in jetty events at $|y|<0.5$ rapidity, which differed from the Monte Carlo predictions. On the other hand, the mean jet multiplicity distributions revealed diverse scaling behavior for the different models, with a better agreement between the \textsc{Hijing++} and \textsc{Pythia 8} calculations. 

\section*{Acknowledgements}
The research was supported by the Hungarian National Research, Development and Innovation Office OTKA K135515, 2019-2.1.11-T\'ET-2019-00078, 2019-2.1.11-T\'ET-2019-00050, 2020-2.1.1-ED-2021-00179, 2019-2.1.6-NEMZ\_KI-2019-00011, 2022-4.1.2-NEMZ\_KI-2022-00008 and 2022-4.1.2-NEMZ\_KI-2022-00009 grants, and by the Wigner Scientific Computing Laboratory and the ELKH Cloud. Author G.B. was supported by the EU project RRF-2.3.1-21-2022-00004 (Artificial Intelligence National Laboratory).

\printbibliography

\end{document}